\documentstyle[12pt]{article}
\newcommand{\bb}{\begin{equation}}
\newcommand{\ee}{\end{equation}}
\newcommand{\bega}{\begin{eqnarray}}
\newcommand{\ega}{\end{eqnarray}}
\newcommand{\begae}{\begin{eqnarray*}}
\newcommand{\egae}{\end{eqnarray*}}

\begin{document}

\centerline{\bf HYDRODYNAMICS OF SPINNING PARTICLES. $^{\dagger}$} 
\footnotetext{$^{\dagger}$ Work partially supported by CAPES, CNPq, and by
INFN, MURST and CNR.}

\vspace*{0.9 cm}

\centerline{Giovanni Salesi$^{(*)}$ \ and \ Erasmo Recami$^{(**)}$} 
\footnotetext{$^{(*)}$ Present address: Dip.to di Fisica, Universit\`a
statale di Catania, 95129--Catania, Italy, \ and \ INFN--Sezione di Catania,
Catania, Italy.} 
\footnotetext{$^{(**)}$ Also: D.M.O./FEEC and C.C.S., UNICAMP, Campinas,
S.P., Brazil. \ Electronic addresses: in Italy, Erasmo.Recami@mi.infn.it; \
in Brazil, Recami@turing.unicamp.br}

\vspace*{0.3 cm}

{\small \centerline{\it Facolt\`a di Ingegneria, Universit\`a Statale di
Bergamo, 24044--Dalmine (BG),} \centerline{\it Italy; \ and \ INFN--Sezione
di Milano, Milan, Italy.} }

\vspace*{0.3 cm}

\centerline{ and }

\vspace*{0.3 cm}

\centerline{Hugo E. Hern\'andez F. \ and \ Luis C. Kretly}

\vspace*{0.3 cm}

{\small \centerline{\it D.M.O./FEEC and C.C.S., State University at
Campinas, Campinas, S.P., Brazil.}}

\vspace*{1.5 cm}

{\bf Abstract \ --} \ In this note, we first obtain the decomposition of the 
non-relativistic field velocity into the classical part (i.e., the velocity 
$\vec{\mbox{\boldmath $w$}}=\vec{\mbox{\boldmath $p$}}/m$  OF the
center-of-mass (CM), and the so-called quantum part (i.e., the velocity 
$\vec{\mbox{\boldmath $V$}}$ of the motion IN the CM frame (namely, the 
internal spin--motion or Zitterbewegung), these two parts being orthogonal.  
Our starting point is 
the Pauli current. \ Then, by inserting such a composite expression of the 
velocity into the kinetic energy term of the non-relativistic newtonian
lagrangian, we get the appearance of the so-called ``quantum potential" 
(which makes the difference between classical and quantum behaviour) as a 
pure consequence of the internal motion. Such a result carries further 
evidence about the possibility that the quantum behaviour of micro-systems 
be a direct consequence of the fundamental existence of spin.

\vspace*{2. cm}

PACS nos.: 11.10.Qr ; \ \ 14.60.Cd ; \ \ 03.70.+k ; \ \ 03.65.-w .\hfill 


\newpage

{\bf 1. \ Madelung fluid: A variational approach}

\hspace*{0.5 cm} The lagrangian for a non-relativistic scalar particle may
be assumed to be: 
\begin{equation}
{\cal L} = \frac{i\hbar}{2}(\psi^{\star}\partial_t\psi -
(\partial_t\psi^{\star})\psi) - \frac{\hbar^2}{2m}{\mbox{\boldmath $\nabla$}}
\psi^{\star} \cdot {\mbox{\boldmath $\nabla$}}\psi - U\psi^{\star}\psi
\end{equation}
where $U$ is the external potential energy and the other symbols have the
usual meaning. \ It is known that, by taking the variations of ${\cal L}$
with respect to $\psi, \; \psi^{\star}$, one can get the Schroedinger
equations for $\psi^{\star}$ and $\psi$, respectively.

\hspace*{0.5 cm} By contrast, since a generic scalar wavefunction $\psi \in$
I$\!\!\!$C can be written as 
\begin{equation}
\psi = \sqrt{\rho} \; {\rm exp} [i \varphi / \hbar] \ ,
\end{equation}
with $\rho,\varphi \in$ I$\!$R, we take the variations of 
\begin{equation}
{\cal L} = -\left[\partial_{t}\varphi + \frac{1}{2m}({
\mbox{\boldmath
$\nabla$}}\varphi)^2 + \frac{\hbar^2}{8m}\left(\frac{{
\mbox{\boldmath
$\nabla$}}\rho}{\rho}\right)^2 + U\right]\rho
\end{equation}
with respect to (w.r.t.) $\rho$ and $\varphi$. \ We then obtain$[1-3]$ the
two equations for the so-called {\it Madelung fluid\/}$[4]$ (which, taken
together, are equivalent to the Schroedinger equation): 
\begin{equation}
\partial_{t}\varphi + \frac{1}{2m}({\mbox{\boldmath $\nabla$}}\varphi)^2 + 
\frac{\hbar^2}{4m}\left[\frac{1}{2}\left(\frac{{\mbox{\boldmath $\nabla$}}
\rho}{\rho}\right)^2 - \frac{\triangle \rho}{\rho}\right] + U = 0
\end{equation}
and 
\begin{equation}
\partial_t \rho + {\mbox{\boldmath $\nabla$}}\cdot (\rho {
\mbox{\boldmath
$\nabla$}}\varphi /m) = 0 \ ,
\end{equation}
which are the Hamilton--Jacobi and the continuity equation for the ``quantum
fluid", respectively; \ where 
\begin{equation}
\frac{\hbar^2}{4m}\left[\frac{1}{2}\left(\frac{{\mbox{\boldmath $\nabla$}}
\rho}{\rho}\right)^2 - \frac{\triangle \rho}{\rho}\right] \equiv -\frac{
\hbar^2}{2m}\frac{\triangle |\psi|}{|\psi|}
\end{equation}
is often called the ``quantum potential". Such a potential derives from the
last-but-one term in the r.h.s. of eq.(3), that is to say, from the (unique)
``non-classical term" 
\begin{equation}
\frac{\hbar^2}{8m}\left(\frac{{\mbox{\boldmath $\nabla$}}\rho}{\rho}\right)^2
\end{equation}
entering our lagrangian ${\cal L}$.

\hspace*{0.5 cm} Notice that we got the present {\it hydrodynamical
reformulation} of the Schroedinger theory {\it directly} from a variational
approach.$[3]$ This procedure, as we are going to see, offers us a physical
interpretation of the non-classical terms appearing in eqs.(3) or (4). \ \
On the contrary, eqs.(4-5) are ordinarily obtained by inserting relation (2)
into the Schroedinger equation, and then separating the real and the
imaginary part: a rather formal procedure, that does not shed light on the
underlying physics.

\hspace*{0.5 cm} Let us recall that an early physical interpretation of the
so-called ``quantum'' potential, that is to say, of term (6) was forwarded
by de Broglie's pilot--wave theory$[5]$; in the fifties, Bohm$[6]$ revisited
and completed de Broglie's approach in a systematic way [and, sometimes,
Bohm's theoretical formalism is referred to as the ``Bohm formulation of
quantum mechanics'', alternative and complementary to Heisenberg's (matrices
and Hilbert spaces), Schroedinger's (wave-functions), and Feynman's (path
integrals) theory]. \ From Bohm's up to our days, several conjectures about
the origin of that mysterious potential have been put forth, by postulating
``sub-quantal'' forces, the presence of an ether, and so on. \ \ Well-known
are also the derivations of the Madelung fluid within the stochastic
mechanics framework:$[7,2]$ in those theories, the origin of the
non-classical term (6) appears as substantially {\it kinematical}. \ In the
non-markovian approaches,$[2]$ for instance, after having assumed the
existence of the so-called zitterbewegung, a spinning particle appears as an
extended-like object, while the ``quantum'' potential is tentatively related
to an internal motion.

\hspace*{0.5 cm} But we do not need following any stochastic approach, even
if our philosophical starting point is the {\it recognition} of the
existence[8-12] of a {\it zitterbewegung} (zbw) or diffusive or {\it 
internal\/} motion [i.e., of a motion observed {\bf in} the center-of-mass
(CM) frame, which is the one where $\mbox{\boldmath $p$}=0$ by definition],
besides of the [external, or drift, or translational, or convective] motion 
{\bf of} the CM. \ In fact, the existence of such an internal motion is
denounced, besides by the mere presence of spin, by the remarkable fact that
in the standard Dirac theory the particle impulse $\mbox{\boldmath $p$}$ is
in general {\it not} parallel to the velocity: \ $\mbox{\boldmath $v$}\neq 
\mbox{\boldmath $p$}/m$; \ moreover, while \ $[\widehat{\mbox{\boldmath $p$}}
,\widehat{H}]=0$ \ so that $\mbox{\boldmath $p$}$ is a conserved quantity,
quantity $\mbox{\boldmath $v$}$ is {\it not} a constant of the motion: \ $[
\widehat{\mbox{\boldmath $v$}},\widehat{H}]\neq 0\ \ (\widehat{
\mbox{\boldmath $v$}}\equiv \mbox{\boldmath $\alpha$}\equiv \gamma ^0
\mbox{\boldmath $\gamma$}$ being the usual vector matrix of Dirac theory). \ 

\hspace*{0.5 cm} For dealing with the zbw it is highly convenient$[10,12]$
to split the motion variables as follows (the dot meaning derivation with
respect to time): 
\begin{equation}
\mbox{\boldmath $x$} = \mbox{\boldmath $\xi$} + \mbox{\boldmath $X$} \; ; \
\ \ \dot{\mbox{\boldmath $x$}} \equiv \mbox{\boldmath $v$} = 
\mbox{\boldmath
$w$} + \mbox{\boldmath $V$} \ ,
\end{equation}
where $\mbox{\boldmath $\xi$}$ and $\mbox{\boldmath $w$} \equiv \dot{
\mbox{\boldmath $\xi$}}$ describe the motion of the CM in the chosen
reference frame, whilst $\mbox{\boldmath $X$}$ and $\mbox{\boldmath $V$}
\equiv \dot{\mbox{\boldmath $X$}}$ describe the internal motion referred to
the CM frame (CMF). \ From a dynamical point of view, the conserved electric
current is associated with the helical trajectories[8-10] of the electric
charge (i.e., with $\mbox{\boldmath $x$}$ and $\mbox{\boldmath $v$} \equiv 
\dot{\mbox{\boldmath $x$}}$), whilst the center of the particle coulombian
field is associated with the geometrical center of such trajectories (i.e.,
with $\mbox{\boldmath $\xi$}$ and $\mbox{\boldmath $w$} \equiv \dot{
\mbox{\boldmath $\xi$}} = \mbox{\boldmath $p$} /m$).

\hspace*{0.5 cm} Going back to lagrangian (3), it is now possible to attempt
an interpretation of the non-classical term ${\frac{\hbar^2}{8m}} ({
\mbox{\boldmath $\nabla$}\rho}/{\rho})^2$ appearing therein. The first term
in the r.h.s. of eq.(3) represents, apart from the sign, the total energy 
\begin{equation}
\partial_t \varphi = - E \ ;
\end{equation}
whereas the second term is recognized to be the kinetic energy $\, 
\mbox{\boldmath $p$}^{2}/2m$ \ {\it of} the CM, if one assumes that 
\begin{equation}
\mbox{\boldmath $p$} = - \mbox{\boldmath $\nabla$} \varphi.
\end{equation}
The third term, that gives origin to the quantum potential, will be shown
below to be interpretable as the kinetic energy {\it in} the CMF, that is,
the internal energy due to the zbw motion. \ It will be soon realized,
therefore, that in lagrangian (3) the sum of the two kinetic energy terms, ${
\mbox{\boldmath $p$}^2}/2m$ and $\frac{1}{2}m\mbox{\boldmath $V$}^2$, is
nothing but a mere application of the K\"onig theorem. \ We are not going to
exploit, as often done, the arrival point, i.e. the Schroedinger equation;
by contrast, we are going to exploit a non-relativistic (NR) analogue of the
Gordon decomposition$[13]$ of the Dirac current: namely, a suitable
decomposition of the {\it Pauli current}.$[14]$ \ In so doing, we shall meet
an interesting relation between zbw and spin.\\

{\bf 2. \ The ``quantum" potential as a consequence of spin and zbw}

\hspace*{0.5 cm} Let us start from the familiar expression of the Pauli
current$[14]$ (i.e., from the Gordon decomposition of the Dirac current in
the NR limit): 
\begin{equation}
\mbox{\boldmath $j$} = \frac{i\hbar}{2m}[(\mbox{\boldmath $\nabla$}
\psi^{\dagger})\psi - \psi^{\dagger} \mbox{\boldmath $\nabla$} \psi] - \frac{
e\mbox{\boldmath $A$}}{m}\psi^{\dagger}\psi + \frac{1}{m}
\mbox{\boldmath
$\nabla$} \wedge (\psi^{\dagger} \widehat{\mbox{\boldmath $s$}}\psi) \ .
\end{equation}
A spinning NR particle can be simply factorized into 
\begin{equation}
\psi \equiv \sqrt{\rho}\> \Phi \ ,
\end{equation}
$\Phi$ being a Pauli 2-component spinor, which has to obey the normalization
constraint 
\[
\Phi^{\dagger}\Phi = 1 
\]
if we want to have $|\psi|^2 = \rho$.

\hspace*{0.5 cm} By definition \ $\rho\mbox{\boldmath $s$} \equiv
\psi^{\dagger}\widehat{\mbox{\boldmath $s$}}\psi \equiv \rho\,\Phi^{\dagger}
\widehat{\mbox{\boldmath $s$}}\Phi$; \ therefore, introducing the
factorization $\psi \equiv \sqrt{\rho}\> \Phi$ into the above expression
(14) for the Pauli current, one obtains: 
\begin{equation}
\mbox{\boldmath $j$} \; \equiv \; \rho \, \mbox{\boldmath $v$} \; = \; \rho
\: \frac{\mbox{\boldmath $p$} - e\mbox{\boldmath $A$}}{m} + \frac{
\mbox{\boldmath $\nabla$}\wedge(\rho\mbox{\boldmath $s$})}{m}
\end{equation}

which is nothing but the decomposition of $\mbox{\boldmath $v$}$ got by
Hestenes$[15]$ by employing the Clifford algebras language: 
\begin{equation}
\mbox{\boldmath $v$} = \frac{\mbox{\boldmath $p$} - e\mbox{\boldmath $A$}}{m}
+ \frac{\mbox{\boldmath $\nabla$} \wedge (\rho\mbox{\boldmath $s$})}{m\rho}
\end{equation}
where the light speed $c$ is assumed equal to 1, quantity $e$ is the
electric charge, $\mbox{\boldmath $A$}$ is the external electromagnetic
vector potential, $\mbox{\boldmath $s$}$ is the {\it spin vector\/} $
\mbox{\boldmath $s$} \equiv \rho^{-1}\psi^{\dagger}\widehat{
\mbox{\boldmath
$s$}}\psi$, and $\widehat{\mbox{\boldmath $s$}}$ is the spin operator
usually represented in terms of Pauli matrices as 
\begin{equation}
\widehat{\mbox{\boldmath $s$}} \equiv \frac{\hbar}{2}(\sigma_{x}; \;
\sigma_{y}; \; \sigma_{z}).
\end{equation}
[Hereafter, every quantity is a {\it local} or {\it field} quantity: ${
\mbox{\boldmath $v$}} \equiv {\mbox{\boldmath $v$}} ({\mbox{\boldmath $x$}}
;t); \ {\mbox{\boldmath $p$}} \equiv {\mbox{\boldmath $p$}} ({
\mbox{\boldmath $x$}} ;t); \ {\mbox{\boldmath $s$}} \equiv {
\mbox{\boldmath
$s$}} ({\mbox{\boldmath $x$}} ;t)$; and so on]. \ \ As a consequence, the
internal (zbw) velocity reads: 
\begin{equation}
\mbox{\boldmath $V$} \equiv \frac{\mbox{\boldmath $\nabla$}\wedge (\rho
\mbox{\boldmath $s$})}{m\rho}.
\end{equation}

\hspace*{0.5 cm} The Schroedinger sub-case [i.e., the case in which the
vector spin field $\mbox{\boldmath $s$}=\mbox{\boldmath $s$}(
\mbox{\boldmath
$x$},t)$ is constant in time and uniform in space] corresponds to {\it spin
eigenstates}; so that one needs now a wave-function factorizable into the
product of a ``non-spin'' part $\sqrt{\rho }e^{i\varphi }$ ({\it scalar\/})
and of a ``{\it spin}'' {\it part} $\chi $: 
\begin{equation}
\psi \equiv \sqrt{\rho }\,e^{i\frac \varphi \hbar }\chi \ ,
\end{equation}
$\chi $ being {\it constant in time and space}. \ Therefore, when $
\mbox{\boldmath $s$}$ has no precession (and no external field is present: $
\mbox{\boldmath $A$}=0$), we have \ $\mbox{\boldmath $s$}\equiv \chi
^{\dagger }\widehat{\mbox{\boldmath $s$}}\chi =$ constant, and 
\begin{equation}
\mbox{\boldmath $V$}=\frac{\mbox{\boldmath $\nabla$}\rho \wedge 
\mbox{\boldmath $s$}}{m\rho }\ne 0\ .\ \ \ \ \ \ \ \ {\rm {(Schroedinger\
case)}}
\end{equation}
One can notice that, {\it even in the Schroedinger theoretical framework,
the zbw does not vanish}, except for plane waves, i.e., for the non-physical
case of $\mbox{\boldmath $p$}$-eigenfunctions, when not only $
\mbox{\boldmath $s$}$, but also $\rho $ is constant and uniform, so that $
\mbox{\boldmath $\nabla$}\rho =0$.

\hspace*{0.5 cm} But let us go on. \ We may now write 
\begin{equation}
\mbox{\boldmath $V$}^2 = \left(\frac{\mbox{\boldmath $\nabla$}\rho\wedge
\mbox{\boldmath $s$}}{m\rho}\right)^2 = \frac{(\mbox{\boldmath $\nabla$}
\rho)^2\mbox{\boldmath $s$}^2 - (\mbox{\boldmath $\nabla$}\rho\cdot
\mbox{\boldmath $s$})^2}{(m\rho)^2}
\end{equation}
since in general it holds 
\begin{equation}
(\mbox{\boldmath $a$} \wedge \mbox{\boldmath $b$})^2 = \mbox{\boldmath $a$}^2
\mbox{\boldmath $b$}^2 - (\mbox{\boldmath $a$}\cdot\mbox{\boldmath $b$})^2 \
.
\end{equation}

Let us now observe that, from the smallness of the negative-energy component
(the so-called ``small component") of the Dirac bispinor it follows the
smallness also of: \ $\mbox{\boldmath $\nabla$}\rho\cdot\mbox{\boldmath $s$}
\simeq 0.$

This was already known from the Clifford algebras approach to Dirac theory,
that yielded$[15]$ ($\beta$ being the Takabayasi angle$[16]$): \ $
\mbox{\boldmath $\nabla$} \cdot (\rho\mbox{\boldmath $s$}) = - m\rho \sin
\beta \;$, \ which in the NR limit corresponds to $\beta = 0$ (``pure
electron'') or $\beta = \pi$ (``pure positron''), so that one gets \ $
\mbox{\boldmath $\nabla$}\cdot (\rho\mbox{\boldmath $s$}) = 0$ \ and in the
Schroedinger case [$\mbox{\boldmath $s$} =$ constant; \ $
\mbox{\boldmath
$\nabla$}\cdot\mbox{\boldmath $s$} = 0$]: 
\begin{equation}
\mbox{\boldmath $\nabla$}\rho\cdot\mbox{\boldmath $s$} = 0.
\end{equation}
By putting such a condition into eq.(19), it assumes the important form 
\begin{equation}
\mbox{\boldmath $V$}^2 = \mbox{\boldmath $s$}^2\left(\frac{{
\mbox{\boldmath
$\nabla$}}\rho}{m\rho}\right)^2 \ ,
\end{equation}
which does {\it finally} allow us to attribute to the so-called
``non-classical" term, eq.(7), of our lagrangian (3) the simple meaning of
kinetic energy of the internal (zbw) motion [i.e., of kinetic energy
associated with the internal (zbw) velocity $\mbox{\boldmath $V$}$],
provided that 
\begin{equation}
\hbar \: = \: 2 \mbox{\boldmath $s$} \ .
\end{equation}
In agreement with the already mentioned K\"onig theorem, such an internal
kinetic energy does appear, in lagrangian (3), as correctly added to the
(external) kinetic energy ${\mbox{\boldmath $p$}}^2 / 2m$ {\bf of} the CM \
[besides to the total energy (9) and the external potential energy $U$].

\hspace*{0.5 cm} Vice-versa, if we assume (within a zbw philosophy) that $
\mbox{\boldmath $V$}$, eq.(22), is the velocity attached to the kinetic
energy term (7), {\it then we can deduce} eq.(23), i.e., we deduce that
actually: 
\[
{|\mbox{\boldmath $s$}|} \: = \: {\frac{1 }{2}} \, \hbar \ . 
\]

\hspace*{0.5 cm} Let us mention, by the way, that in the stochastic
approaches the (``non-classical") stochastic, diffusion velocity is $
\mbox{\boldmath $V$} \equiv \mbox{\boldmath $v$}_{{\rm dif}} = \nu \, ({{
\mbox{\boldmath $\nabla$} \rho} / \rho})$, quantity $\nu$ being the
diffusion coefficient of the ``quantum" medium. \ In those approaches,
however, one has to {\it postulate} that \ $\nu \equiv \hbar / 2m$. \ In our
approach, on the contrary, if we just adopted for a moment the stochastic
language, by comparison of our eqs.(7), (22) and (23) we would immediately 
{\it deduce} that $\nu = \hbar / 2m$ and therefore the interesting relation 
\begin{equation}
\nu = \frac{|\mbox{\boldmath $s$}|}{m} \ .
\end{equation}

\hspace*{0.5 cm} Let us explicitly remark that, because of eq.(22), in the
Madelung fluid equation (and therefore in the Schroedinger equation)
quantity $\hbar$ is naturally replaced by $2|\mbox{\boldmath $s$}|$, the
presence itself of the former quantity being no longer needed; we might say
that it is more appropriate to write $\hbar = 2|\mbox{\boldmath $s$}|$,
rather than\break 
$|\mbox{\boldmath $s$}| = \hbar /2$ \ldots!

\hspace*{0.5 cm} We conclude by stressing the following. \ We first achieved
a non-relativistic, Gordon-like decomposition of the field velocity within
the ordinary tensorial language. \ Secondly, we derived the ``quantum"
potential (without the postulates and assumptions of stochastic quantum
mechanics) by simply relating the ``non-classical" energy term to zbw and
spin. \ Such results carry further evidence that the quantum behaviour of
micro-systems may be a direct consequence of the existence of spin. In fact,
when $\mbox{\boldmath $s$} = 0$, the quantum potential does vanish in the
Hamilton--Jacobi equation, which then becomes a totally {\it classical} and
newtonian equation. \ We have also seen that quantity $\hbar$ itself enters
the Schroedinger equation owing to the presence of spin. \ We are easily
induced to conjecture that no scalar {\it quantum} particles exist that are
really elementary; \ but that scalar particles are always constituted by
spinning objects endowed with zbw.

\hspace*{0.5 cm} The authors wish to acknowledge stimulating discussions
with R.J.S. Chisholm, J. Keller, Z. Oziewicz and J. Vaz. \ For the kind
cooperation, thanks are also due to Alzetta, I. Arag\'on, A.P.L. Barbero, A.
Bonasera, E.C. Bortolucci, R. Garattini, G. Giuffrida, C. Kiihl, G.
Marchesini, R.L. Monaco, E.C. Oliveira, Peruzzi, R. Petronzio, M.
Pignanelli, R.M. Salesi, S. Sambataro, M. Santini, D.S. Thober, E. Tonti, I.
Torres Lima Jr., M.T. Vasconselos and particularly to L. Bosi, A. Gigli
Berzolari, G.C. Cavalleri e C. Dipietro. \ This work is dedicated to the
memory of Asim O. Barut.


\begin{thebibliography}{16}
\bibitem{1}  F. Guerra and L.M. Morato: {\it Phys. Rev. D\/}{\bf 27} (1983)
1774.

\bibitem{2}  G. Cavalleri et al.: {\it Lett. Nuovo Cim.} {\bf 43} (1985)
285; \ {\it Nuovo Cim. B\/}{\bf 95} (1986) 194; \ {\it Phys. Rev. B\/}{\bf 41
} (1990) 6751; \ {\it Phys. Rev. B\/}{\bf 43} (1991) 3223. \ Cf. also G.
Cavalleri and G. Salesi: ``$\hbar $ derived from cosmology and origin of
special relativity and QED,'' to appear in the {\it Proceedings of
``Physical Interpretations of Relativity Theory (British Society for the
Philosophy of Science; London, 9--12 September, 1994)''}.

\bibitem{3}  Salesi G.: ``Spin and Madelung fluid'', in press.

\bibitem{4}  E. Madelung: {\it Z. Phys.} {\bf 40} (1926) 332. \ Cf. also
Rylov Y.A.: {\it Adv. Appl. Cliff. Alg.} {\bf 5} (1995) 1; \ T. Waite: ``The
relativistic Helmholtz theorem and solitons'', {\it Phys. Essays} {\bf 8}
(1995) 60.

\bibitem{5}  L. de Broglie: {\it C. R. Acad. Sc. (Paris)} {\bf 183} (1926)
447; \ {\it Non-Linear Wave Mechanics} (Elsevier; Amsterdam, 1960).

\bibitem{6}  D. Bohm: {\it Phys. Rev.} {\bf 85} (1952) 166; \ {\bf 85}
(1952) 180; \ D. Bohm and J.P. Vigier: {\it Phys. Rev.} {\bf 96} (1954) 208.

\bibitem{7}  G.C. Ghirardi, C. Omero, A. Rimini and T. Weber: {\it Rivista
Nuovo Cimento} {\bf 1} (1978) 1, and refs. therein; \ Guerra F.: {\it Phys.
Rep.} {\bf 77} (1981) 263, and refs. therein.

\bibitem{8}  A.H. Compton: {\it Phys. Rev.} {\bf 14} (1919) 20, 247, and
refs. therein. \ See also Bostick W.H.: ``Hydromagnetic model of an
elementary particle'', in {\it Gravity Res. Found. Essay Contest} (1958 and
1961).

\bibitem{9}  E. Schroedinger: {\it Sitzunger. Preuss. Akad. Wiss. Phys.
Math. Kl.} {\bf 24} (1930) 418; {\bf 3} (1931) 1. \ Cf. also P.A.M. Dirac: 
{\it The Principles of Quantum Mechanics} (Claredon; Oxford, 1958), $4^{{\rm 
th}}$ edition, p. 262; \ J. Maddox J.: ``Where Zitterbewegung may lead'', 
{\it Nature} {\bf 325} (1987) 306; \ E. Gonz\'alez, H. E. Hern\'andez-Figueroa
and F. A. Fern\'andez: {\it IEEE Trans. on Magn.} {\bf 31} (1995) 1741; \
H. E. Hern\'andez-Figueroa and M. L. Brandao: {\it IEEE Photon. Techn. Lett.} 
{\bf 9} (1997) 351.

\bibitem{10}  {\it a\/}) A.O. Barut et al.: {\it Phys. Rev. Lett.} {\bf 52}
(1984) 2009; \ {\it Phys. Rev. D\/}{\bf 23} (1981) 2454; {\it D\/}{\bf 24}
(1981) 3333; \ {\it Phys. Rev. Lett.} {\bf 53} (1984) 2355; \ {\it Phys.
Lett. B\/}{\bf 237} (1990) 436; \ {\it Phys. Lett. A\/}{\bf 189} (1994) 277;
\ \ {\it b\/}) A.O. Barut and M. Pav\v {s}i\v {c}: {\it Class. Quant. Grav.} 
{\bf 4} (1987) L131; \ {\it Phys. Lett. B\/}{\bf 216} (1989) 297; \ M. Pav\v
{s}i\v {c}: {\it Phys. Lett. B\/}{\bf 205} (1988) 231; {\it B\/}{\bf 221}
(1989) 264; \ {\it Class. Quant. Grav.} {\bf 7} (1990) L187; \ and refs.
therein.

\bibitem{11}  H.C. Corben: {\it Classical and Quantum Theories of Spinning
Particles} (Holden-Day; San Francisco, 1968); \ {\it Phys. Rev.} {\bf 121}
(1961) 1833; \ {\it Phys. Rev. D\/}{\bf 30} (1984) 2683; \ {\it Am. J. Phys.}
{\bf 45} (1977) 658; {\bf 61} (1993) 551; \ {\it Int. J. Theor. Phys.} {\bf 
34} (1995) 19, and refs. therein.

\bibitem{12}  M. Pav\v {s}i\v {c}, E. Recami, W.A. Rodrigues, G.D.
Maccarrone, F. Raciti and G. Salesi: {\it Phys. Lett. B\/}{\bf 318} (1993)
481; \ W.A. Rodrigues, J. Vaz, E. Recami and G. Salesi: {\it Phys. Lett. B\/}
{\bf 318} (1993) 623; \ G. Salesi and E. Recami: {\it Phys. Lett. A\/}{\bf 
190} (1994) 137, E389; \ ``Field theory of the extended-like electron'', in 
{\it Particles, Gravity and Space-Time}, ed. by P.I. Pronin \& G.A.
Sardanashvily (World Scient.; Singapore, 1996), pp.345-368; \ ``About the
kinematics of spinning particles'', submitted for pub.; \ E. Recami and G.
Salesi: {\it Adv. Appl. Cliff. Alg.} {\bf 6} (1996) 27-36; \ and refs.
therein.

\bibitem{13}  J.D. Bjorken and S.D. Drell: {\it Relativistic Quantum
Mechanics}, p.36 (McGraw--Hill; U.S.A., 1964).

\bibitem{14}  Landau L.D. and E.M. Lifshitz: {\it Relativistic Quantum Theory
} (Addison-Wesley; Reading, Mass.).

\bibitem{15}  D. Hestenes: {\it Space-Time Algebra} (Gordon \& Breach; New
York, 1966); \ {\it New Foundations for Classical Mechanics} (Kluwer;
Dordrecht, 1986); \ {\it Found. Phys.} {\bf 23} (1993) 365; {\bf 20} (1990)
1213; {\bf 15} (1985) 63; {\bf 12} (1981) 153; \ {\it Am. J. Phys.} {\bf 47}
(1979) 399; {\bf 39} (1971) 1028; {\bf 39} (1971) 1013; \ {\it J. Math. Phys.
} {\bf 14} (1973) 893; {\bf 16} (1975) 556; {\bf 16} (1975) 573; {\bf 8}
(1967) 798; {\bf 8} (1967) 809; {\bf 8} (1967) 1046; \ D. Hestenes and G.
Sobczyk: {\it Clifford Algebra to Geometric Calculus} (Reidel; Dordrecht,
1984).

\bibitem{16}  T. Takabayasi: {\it Nuovo Cim.} {\bf 3} (1956) 233; \ {\bf 7}
(1958) 118; \ T. Takabayasi and J.-P. Vigier: {\it Progr. Theor. Phys.} {\bf 
18} (1957) 573.
\end{thebibliography}
\end{document}